\documentclass[aip,jap,reprint,showpacs,superscriptaddress]{revtex4-1}

\usepackage{graphicx}
\usepackage{dcolumn}
\usepackage{bm}
\usepackage{amsmath,amssymb}
\usepackage{comment}

\def\U#1{{%
\def\O{\mbox{O}}
\def\u{\mbox{u}}
\mathcode`\u=\mu
\mathcode`\O=\Omega
\mathrm{#1}}}

\def\Re{\mathop{\mathrm{Re}}}
\def\Im{\mathop{\mathrm{Im}}}

\def\ii{{\mathrm{i}}}
\def\dd{{\mathrm{d}}}
\def\sub#1{_{\scriptsize\mbox{#1}}}

\begin{document}


\title{Microplasma generation by slow microwave in an
electromagnetically induced transparency-like metasurface} 

\author{Yasuhiro Tamayama}
\email{tamayama@vos.nagaokaut.ac.jp}
\affiliation{Department of Electrical, Electronics and
Information Engineering, Nagaoka University of
Technology, 1603-1 Kamitomioka, Nagaoka, Niigata 940-2188, Japan}
\author{Osamu Sakai}
\affiliation{Electronic Systems Engineering, The University of Shiga
Prefecture, 2500 Hassakacho, Hikone, Shiga 522-8533, Japan}
\date{\today}

\begin{abstract}
Microplasma generation using microwaves in an electromagnetically induced transparency (EIT)-like metasurface composed of two types of radiatively coupled cut-wire resonators with slightly different resonance frequencies is investigated. Microplasma is generated in either of the gaps of the cut-wire resonators as a result of strong enhancement of the local electric field associated with resonance and slow microwave effect. The threshold microwave power for plasma ignition is found to reach a minimum at the EIT-like transmission peak frequency, where the group index is maximized. A pump--probe measurement of the metasurface reveals that the transmission properties can be significantly varied by varying the properties of the generated microplasma near the EIT-like transmission peak frequency and the resonance frequency. The electron density of the microplasma is roughly estimated to be of order $1\times 10^{10}\,\U{cm}^{-3}$ for a pump power of 15.8\,W by comparing the measured transmission spectrum for the probe wave with the numerically calculated spectrum. In the calculation, we assumed that the plasma is uniformly generated in the resonator gap, that the electron temperature is 2\,eV, and that the elastic scattering cross section is $20 \times 10^{-16}\,\U{cm}^2$.
\end{abstract}

\maketitle

\section{Introduction}

Plasma generation using high-frequency electromagnetic waves is a key issue not only in scientific research but also for industrial applications in various inorganic and organic materials processing methods. Microwaves at a few gigahertz with a high power of up to several hundreds of watts are practical energy sources from an economical point of view. Conventional microwave plasma sources guide microwaves to the plasma region in propagation modes such as typical waveguide and surface wave modes; this is partly because material processing requires the uniform treatment of wafers larger than several centimeters.

In recent decades, microplasmas smaller than several millimeters have attracted significant attention\cite{iza08,tachibana10} because their electron density is fairly high and suitable for the treatment of biomedical materials\cite{fridman13} and the creation of nanoparticles.\cite{mariotti10} Some microwave sources that have been proposed for microplasma
generation are based on intensified microwaves in resonant structures, such as split-ring resonators,\cite{pendry99} whose gap regions work as capacitors before plasma generation.\cite{iza05,kim05,singh-p_14}

In a recent study, our group demonstrated plasma ignition in an electromagnetically induced transparency (EIT)-like metasurface.\cite{tamayama15_prb} EIT is a quantum interference phenomenon that occurs in atoms interacting with electromagnetic fields,\cite{hau99,fleischhauer05} and various types of metasurfaces and metamaterials that mimic EIT have been intensively studied.\cite{fedotov07,zhang_prl08,tassin_prl09,liu_nat09,tamayama10,kurter11,tamayama12,gu12,nakanishi13,miyamaru14,moritake15} An important feature of this EIT-like metasurface, which is composed of two types of radiatively coupled cut-wire resonators, is the local electric field enhancement, which is stronger than that in a metasurface composed of only one type of cut-wire resonator (Lorentz-type metasurface) because of the compression of the electromagnetic energy density associated with the low-group-velocity propagation. This implies that studying EIT-like metasurfaces and metamaterials may lead to the development of a useful method for plasma generation. Although plasma ignition in the EIT-like metasurface was demonstrated in our previous study, only the differences between the transmission characteristics of the metasurface with and without the plasma were discussed. It has not been verified that the local electric field enhancement factor increases with decreasing group velocity. Also, the electron density of the generated plasma has not been evaluated. These issues should be clarified within the context of plasma physics, electromagnetic optics, and other related fields of study.

In this study, the dependence of the threshold incident power for plasma ignition in the EIT-like metasurface and the characteristics of the generated plasma on various parameters is investigated by measuring the transmittance of the metasurface to clarify the above issues. The frequency dependence of the threshold power for plasma ignition reveals that the local electric field enhancement is strongest when the group velocity is lowest. The pressure dependence of the transmission spectrum of this metasurface with microplasma provides an insight into the nature of the generated plasmas. The theory for enhancing a local electric field used in this study can be applied to efficient generation of various nonlinear phenomena as well as low-power ignition of plasma. In addition, the operating frequency of metasurfaces/metamaterials is scalable by varying the size of unit cell. Therefore, efficient generation of plasma with any desired size would be achieved using this theory.

\section{microplasma generation in EIT-like metasurface}

\begin{figure}[tb]
 \begin{center}
  \includegraphics[scale=0.85]{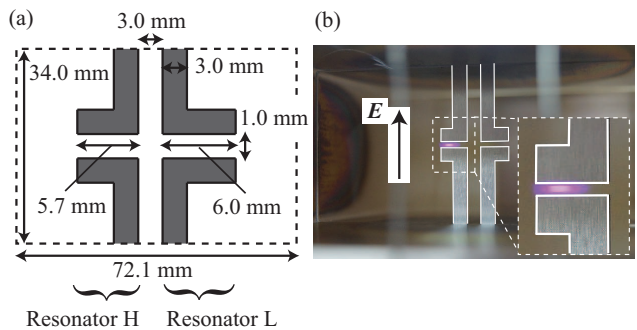}
  \caption{(a) Unit structure of the EIT-like metasurface.
  The metasurface is made of aluminum with a thickness of
  $1.0\,\U{mm}$. (b) Photograph of the metasurface when microplasma is generated
  in the gap of resonator H. The outline of the structure is marked by white solid lines.
  The inset shows the magnification of the gap of resonator H.}
  \label{fig:metasurface}
 \end{center}
\end{figure}

This section briefly describes the theory behind the generation of microplasma in the EIT-like metasurface developed in our previous study.\cite{tamayama15_prb} The unit structure of the EIT-like metasurface used in this study is shown in Fig.~\ref{fig:metasurface}(a). The metasurface is composed of two types of cut-wire resonators with slightly different resonance frequencies. The resonance frequency of the cut-wire resonator is determined by the inductance of the metallic pattern and the capacitance of the gap. The capacitance of the cut-wire resonator on the left-hand side of Fig.~\ref{fig:metasurface}(a) is smaller than that on the right-hand side; thus, the resonance frequency of the cut-wire resonator on the left-hand side is higher than that on the right-hand side. Hereafter, the resonators on the left- and right-hand sides of Fig.~\ref{fig:metasurface}(a) are referred to as resonators H and L, respectively. These two resonators are arranged such that they are coupled via a radiation mode.\cite{verslegers12,zhang_s_12} The distance between the two resonators is tuned such that the direct coupling between the resonators vanishes.\cite{tamayama14} Assuming that the electric susceptibility $\chi\sub{e}$ of the metasurface is proportional to the sum of the charges $q\sub{H}$ and $q\sub{L}$ stored in the gaps of resonators H and L, the electric susceptibility is given by
\begin{equation}
\chi\sub{e}
\approx
\frac{-\alpha ( \omega^2 - \omega_0^2 + \ii \gamma^{\prime} \omega)}{
(\omega^2 - \omega_0^2 + \ii \gamma_0 \omega)^2
- [(\omega\sub{H}^2 - \omega\sub{L}^2)/2]^2 + (\gamma\sub{M} \omega )^2 }
, \label{eq:40}
\end{equation}
where $\omega$ is the angular frequency of the incident electromagnetic wave; $\omega\sub{H}$ and $\omega\sub{L}$ are the resonance angular frequencies of resonators H and L, respectively; $\gamma_0$ represents the sum of the radiative and nonradiative losses of each resonator; $\gamma\sub{M}$ the radiative coupling; $\gamma^{\prime}= \gamma_0 - \gamma\sub{M}$ the power dissipated in each resonator that does not contribute to the radiative coupling; $\alpha$ is a proportionality constant; and $\omega_0 = \sqrt{(\omega\sub{H}^2 +\omega\sub{L}^2)/2} \approx (\omega\sub{H} + \omega\sub{L})/2$.\cite{tamayama14} Equation \eqref{eq:40} is similar to the electric susceptibility for EIT.\cite{hau99,fleischhauer05,alzar02} A narrowband transparency window appears in a broad absorption line. At the center of this transparency window ($\omega = \omega_0$), the value of $\dd \Re{(\chi\sub{e})}/ \dd \omega$ is largest, and hence the group index reaches a maximum. A strongly enhanced electric field is induced in the gaps of the cut-wire resonators owing to the enhancement of the electromagnetic energy density associated with a large group index.\cite{krauss08,chen10_pra} Assuming that $\gamma^{\prime} \ll \gamma_0$ is satisfied, which implies that the radiative coupling between the two resonators is strong, the group index and local electric field enhancement at $\omega = \omega_0$ are maximized when  $\omega\sub{H} - \omega\sub{L} \approx \sqrt{2\gamma_0\gamma^{\prime}}$. The geometrical parameters shown in Fig.~\ref{fig:metasurface} are defined such that this condition is satisfied. In this case,
$\omega\sub{H} = 2\pi \times 3.059\,\U{GHz}$,
$\omega\sub{L} = 2\pi \times 3.016\,\U{GHz}$,
$\gamma_0 = 2\pi \times 920\,\U{MHz}$, and
$\gamma^{\prime} = 2\pi \times 0.966\,\U{MHz}$.
At $\omega = \omega_0 = 2\pi \times 3.031\,\U{GHz}$, the group delay is $91\,\U{ns}$, which corresponds to a group index of $2.7\times 10^4$, and a local electric field with an amplitude approximately 300 times larger than that of the incident electromagnetic wave is induced in the gaps of the resonators.

Figure \ref{fig:metasurface}(b) shows a photograph of the EIT-like metasurface in a rectangular waveguide with cross-sectional dimensions of 34.0\,mm $\times$ 72.1\,mm when a continuous microwave of a certain power and an angular frequency $\omega_0$ is incident on the metasurface. (Because the walls of the rectangular waveguide behave almost as periodic boundaries, a periodically arranged array of the unit cell is equivalently realized by placing only a single unit cell in the waveguide. Therefore, the fabricated structure can be safely referred to as a metasurface.) A microplasma is generated in the gap of resonator H because of the strongly enhanced local electric field. As described in Sec.\,\ref{sec:result1}, the incident frequency determines the gap in which the microplasma is generated.

\section{Experimental methods}

\begin{figure}[tb]
 \begin{center}
  \includegraphics[scale=0.5]{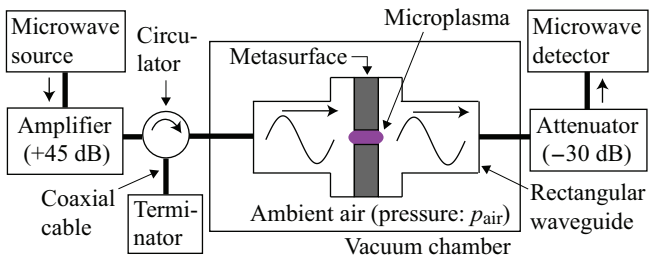}
  \caption{Schematic of experimental setup.
  See main text for details regarding the microwave source and microwave detector.}
  \label{fig:setup1}
 \end{center}
\end{figure}

Using the experimental setup schematically shown in Fig.~\ref{fig:setup1}, we investigated the threshold incident power for plasma ignition in the EIT-like metasurface and the characteristics of the generated plasma. The EIT-like metasurface was placed in a rectangular waveguide with cross-sectional dimensions of 34.0\,mm $\times$ 72.1\,mm. The waveguide was placed in an acrylic vacuum chamber where the pressure of the ambient air was reduced to $p\sub{air}$.

In evaluating the threshold incident power for plasma ignition, a signal generator and a spectrum analyzer were used as microwave source and microwave detector, respectively (Fig.~\ref{fig:setup1}). The power of the continuous microwave incident on the metasurface was assumed to be equal to the output power $P_0$ of the signal generator multiplied by the gain of the amplifier, which was $10^{4.5}$. The dependence of the transmittance on incident power was measured by sweeping the incident power up and down. Here, the transmittance represents the power detected by the spectrum analyzer divided by $P_0$.

In examining the characteristics of the generated plasma, a signal generator and a network analyzer were used as microwave sources. The signal generator generated a high-power microwave for plasma generation, and the network analyzer generated a low-power microwave (power incident on the metasurface: 0.16\,W) for the evaluation of the transmittance of the metasurface. The former and latter microwaves are referred to as the pump wave and probe wave, respectively. These two waves were combined, amplified, and incident on the metasurface. The pump power incident on the metasurface was assumed to be $10^{4.2}P_0$, as determined from the gain of the amplifier ($+45\,\U{dB}$) and the transmittance of the combiner ($-3\,\U{dB}$). The pump frequency was fixed at 3.031\,GHz ($=\omega_0 / 2\pi$). Under this condition, microplasma can be generated in the gap of resonator H, as described above and in Sec.\,\ref{sec:result1}. The transmitted probe wave was detected by the network analyzer to measure the transmittance, defined as the total transmittance from the output port of the network analyzer to its input port in this experiment.

\section{dependence of threshold power for plasma ignition on incident frequency and gas pressure}\label{sec:result1}

\begin{figure}[tb]
 \begin{center}
  \includegraphics[scale=0.75]{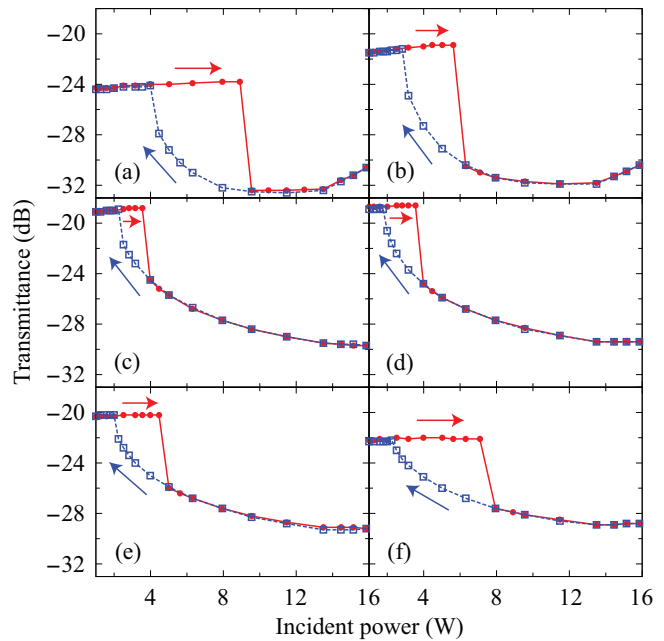}
  \caption{Power dependence of transmittance at
  $p\sub{air} = 1.0\,\U{kPa}$ and incident frequencies of
  (a) 3.028\,GHz, (b) 3.029\,GHz, (c) 3.030\,GHz,
  (d) 3.031\,GHz, (e) 3.032\,GHz, and (f) 3.033\,GHz.
  The solid circles and open squares represent transmittance when the incident power
  is swept up and down, respectively. Lines are included purely to act as visual guides.}
  \label{fig:freq}
 \end{center}
\end{figure}

First, we present the measured results related to the investigation of the threshold power for plasma ignition in the EIT-like metasurface. Figure \ref{fig:freq} shows the incident power dependence of the transmittance of the metasurface for $p\sub{air} = 1.0\,\U{kPa}$ at six different incident frequencies. For every case, as the incident power increases, transmittance first remains unchanged, discontinuously drops when it exceeds a certain value, and then continuously varies. The ignition of the microplasma occurs in the gap of one of the cut-wire resonators when transmittance diminishes discontinuously. This is because the generation of microplasma causes the suppression of the narrowband transparency window, to be discussed later and shown in Fig.~\ref{fig:spec_3dBm}. The threshold power for plasma ignition is lowest when the incident frequency is 3.030\,GHz and 3.031\,GHz. The threshold increases as the incident frequency varies from the transmission peak frequency of 3.031\,GHz. (The linear transmission characteristics without plasma and minimum threshold power are slightly different from those obtained in our previous study.\cite{tamayama15_prb} This is the result of a slight difference in position of the metasurface in the waveguide.) This implies that the local electric field enhancement is strongest at the transmission peak frequency, at which the group index is largest. The microplasma generation was observed immediately near $\omega=\omega_0$ and was not observed near $\omega = \omega\sub{L}$ or $\omega\sub{H}$ in this experiment, where the maximum power incident on the metasurface was $16\,\U{W}$. This is because the local electric field enhancement factor at $\omega = \omega\sub{L}$ and $\omega\sub{H}$ is equal to that of the metasurface composed of one type of cut-wire resonator. The ratio of the enhancement factor at $\omega = \omega_0$ to that at $\omega = \omega\sub{L}$ and $\omega\sub{H}$ is approximately $q\sub{L} (\omega_0) / q\sub{L} (\omega\sub{L}) \approx q\sub{H} (\omega_0) / q\sub{H} (\omega\sub{H}) \approx \gamma_0 /[2(\omega\sub{H} -\omega\sub{L})] = 11$. Therefore, microplasma can also be generated near $\omega = \omega\sub{L}$ and $\omega\sub{H}$ for an incident power higher than $11^2 \times 4 \,\U{W} = 5 \times 10^2 \,\U{W}$. Additionally, microplasmas can be generated simultaneously in both gaps at around $\omega = \omega_0$ for such a high-power microwave.

There exists an incident power regime in which the transmittance is
different when the power is swept up and when swept down. That is, the power dependence has clear hysteretic properties. This arises from memory effects resulting from residual charged particles, i.e., electrons and ions. Before plasma ignition, electron-impact ionization is very difficult because there is a lack of initial electrons, and the threshold for ignition is high enough to accelerate a very small amount of electrons to ionize gases. In contrast, when plasmas exist and sufficient electrons and ions remain in a given space, the plasma state is easily sustained. Consequently, the minimum power required to sustain plasma is different from the threshold power for plasma ignition, and two values of the transmittance exist for one input microwave power when the incident power is between these two powers. In general, such memory or time-lag effects are essential for nonlinear dynamics.

The power dependence of transmittance at incident frequencies 3.028\,GHz and 3.029\,GHz differs from that at incident frequencies ranging from 3.030\,GHz to 3.033\,GHz. This implies that microplasma is generated in resonators L and H in the former and latter frequency regions, respectively. This observation can be understood from the fact that plasma ignition occurs in the cut-wire resonator with the higher local electric field enhancement factor. Equation \eqref{eq:40} demonstrates that $q\sub{L} \simeq -q\sub{H}$ is satisfied at the transmission peak frequency. Because the capacitance of resonator H is smaller than that of resonator L, the local electric field enhancement in resonator H is stronger than that in resonator L at the transmission peak frequency; therefore, plasma ignition occurs in resonator H. As the incident frequency increases (decreases) from the transmission peak frequency, $| q\sub{H} | / | q\sub{L} |$ increases (decreases), and the ratio of the local electric field enhancement factor in resonator H to that in resonator L increases (decreases) because the incident frequency approaches the resonance frequency of resonator H (resonator L). When the ratio of these field enhancement factors is larger (smaller) than 1, plasma ignition occurs in resonator H (resonator L). It has already been demonstrated that the transmittance of the metasurface depends on the resonator in which the microplasma is generated;\cite{tamayama15_prb} thus, plasma ignition occurs in resonator L (resonator H) when the incident frequency is less (greater) than $f\sub{e}$, which is between 3.029\,GHz and 3.030\,GHz.

\begin{figure}[tb]
 \begin{center}
  \includegraphics[scale=1]{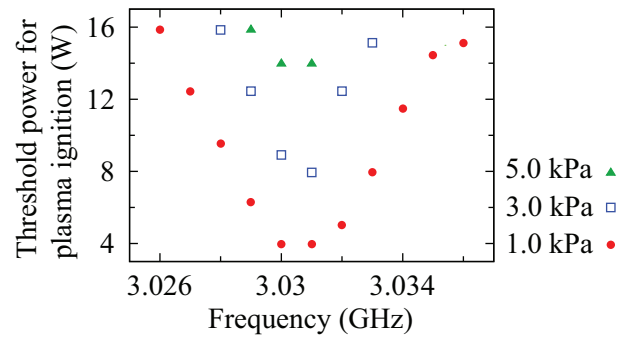}
  \caption{Threshold power for plasma ignition as a function of incident
  frequency for $p\sub{air}=1.0\,\U{kPa}$, 3.0\,kPa, and 5.0\,kPa. }
  \label{fig:thre}
 \end{center}
\end{figure}

Figure \ref{fig:thre} shows the frequency dependence of the threshold power for plasma ignition at $p\sub{air}=1.0\,\U{kPa}$, 3.0\,kPa, and 5.0\,kPa. The threshold value is smallest at the transmission peak frequency at every value of $p\sub{air}$; thus, the strong enhancement of the local electric field resulting from the large group index is experimentally confirmed. As the pressure is raised from 1.0\,kPa to 5.0\,kPa, the threshold power for ignition increases. This dependence is consistent with the Paschen curve. As described in Ref.~\onlinecite{engel_55}, the ignition voltage for direct current operations reaches a minimum at $pd \approx 1\,\U{Torr}\cdot\U{cm}$ for air, where $p$ is the gas pressure and $d$ is the electrode distance. Usually, as the frequency of the applied voltage increases, this minimum $pd$ condition shifts toward lower pressures. That is, the conditions of the discharge space regulated in the present experiment lies on the right-hand side of the Paschen curve, where the ignition voltage increases with increasing $p$.

\section{dependence of microplasma properties on pump power and gas pressure}

\begin{figure}[tb]
 \begin{center}
  \includegraphics[scale=1]{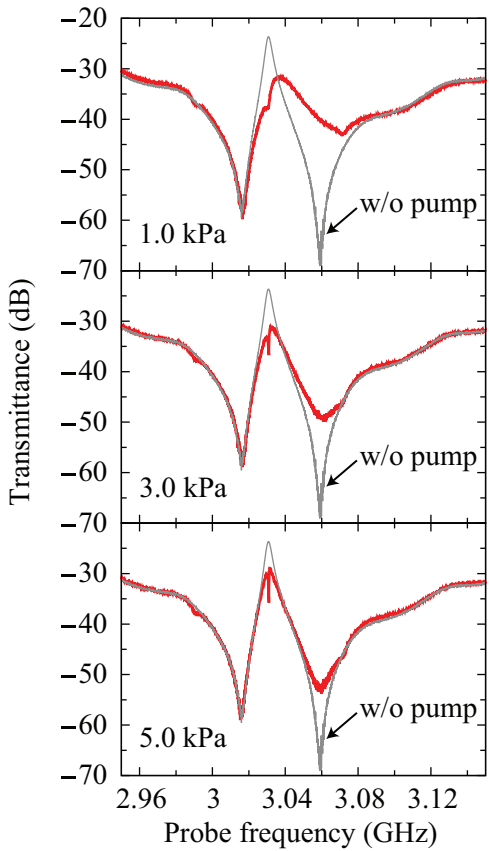}
  \caption{Transmission spectra for the probe wave at an incident pump power of
  15.8\,W and $p\sub{air} = 1.0\,\U{kPa}$, 3.0\,kPa, and 5.0\,kPa. For comparison,
  the transmission spectrum without the pump wave is also shown as the thin light gray curve.}
  \label{fig:spec_3dBm}
 \end{center}
\end{figure}

Next, we describe the results related to the investigation of the plasma properties. Figure \ref{fig:spec_3dBm} shows the transmission spectra of the metasurface for the probe wave at a pump power of 15.8\,W and $p\sub{air} =1.0\,\U{kPa}$, 3.0\,kPa, and 5.0\,kPa. The higher-frequency transmission dip becomes shallower and shifts to slightly higher frequencies compared with that without the pump wave. This is because the transmission dips correspond to the individual resonances of the two types of cut-wire resonators~\cite{tamayama14} and the real and imaginary parts of the effective permittivity in the gap of resonator H decrease and increase, respectively, as a result of the generation of microplasma. The increase in the imaginary part also causes an increase in $\gamma^{\prime}$, which corresponds to a decrease in the transmittance at the transmission peak frequency. This indicates that transmittances near the resonance frequency and the transmission peak frequency can be significantly varied by varying the properties of the generated microplasma.

\begin{figure}[tb]
 \begin{center}
  \includegraphics[scale=1]{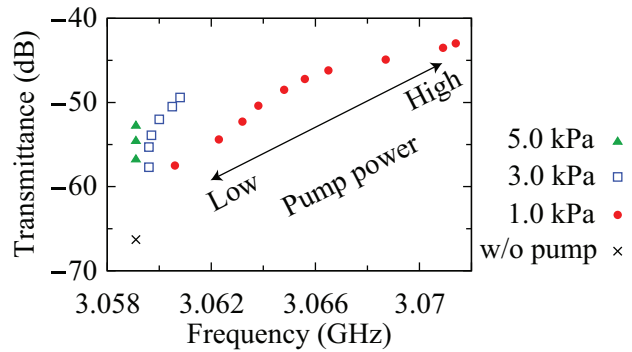}
  \caption{Relationship between the center frequency of
  the higher-frequency transmission dip and the transmittance
  at that frequency for $p\sub{air} =1.0\,\U{kPa}$, 3.0\,kPa, and 5.0\,kPa.
  The data points correspond to those in Fig.~\ref{fig:dens};
  see Fig.~\ref{fig:dens} for the pump power at each point. }
  \label{fig:dip}
 \end{center}
\end{figure}

To evaluate the parameter dependence of the permittivity of the generated microplasma, the transmission spectrum of the metasurface was measured while varying the incident pump power from 15.8\,W to lower powers. Figure \ref{fig:dip} shows the relationship between the center frequency of the higher-frequency transmission dip and the transmittance at that frequency at $p\sub{air} = 1.0\,\U{kPa}$, 3.0\,kPa, and 5.0\,kPa. As the pump power increases, the shift to higher frequencies becomes larger, and the transmission dip becomes shallower. This implies that the real and imaginary parts of the permittivity decrease and increase, respectively, with increasing the pump power. Additionally, as $p\sub{air}$ increases, the ratio of the transmittance variation to the frequency shift increases.

The above results can be understood by considering the dependence of the permittivity of plasma on the gas pressure and electron density. The relative permittivity of plasma is given by the Drude dispersion relation:
\begin{equation}
  \varepsilon\sub{r} = 1- \frac{\omega\sub{p}^2}{\omega ( \omega + \ii
  \gamma)}, \label{eq:drude}
\end{equation}
where $\omega\sub{p} = \sqrt{(e^2 n\sub{e})/ (m\sub{e} \varepsilon_0 )}$ is the plasma angular frequency, $\gamma = \sigma n\sub{N} \sqrt{(8k\sub{B} T\sub{e})/(\pi m\sub{e})}$ the elastic-collision angular frequency, $e$ the electron charge, $n\sub{e}$ the electron density, $m\sub{e}$ the electron mass, $\varepsilon_0$ the permittivity in a vacuum, $\sigma$ the elastic scattering cross section, $n\sub{N}$ the neutral particle density, $k\sub{B}$ Boltzmann's constant, and $T\sub{e}$ the electron temperature. Equation \eqref{eq:drude} is reduced to $\Im{(\varepsilon\sub{r})}/[1-\Re{(\varepsilon\sub{r})}] = \gamma / \omega$. Because $\gamma$ is proportional to $n\sub{N}$, the ratio of $\Im{(\varepsilon\sub{r})}$ to $[1-\Re{(\varepsilon\sub{r})}]$ increases with increasing $p\sub{air}$.\cite{sakai12_psst} The numerator and denominator of this ratio correspond to the transmittance variation and frequency shift of the transmission dip, respectively; therefore, this observation qualitatively agrees with the results shown in Fig.~\ref{fig:dip}.

\begin{figure*}[tb]
 \begin{center}
  \includegraphics[scale=0.57]{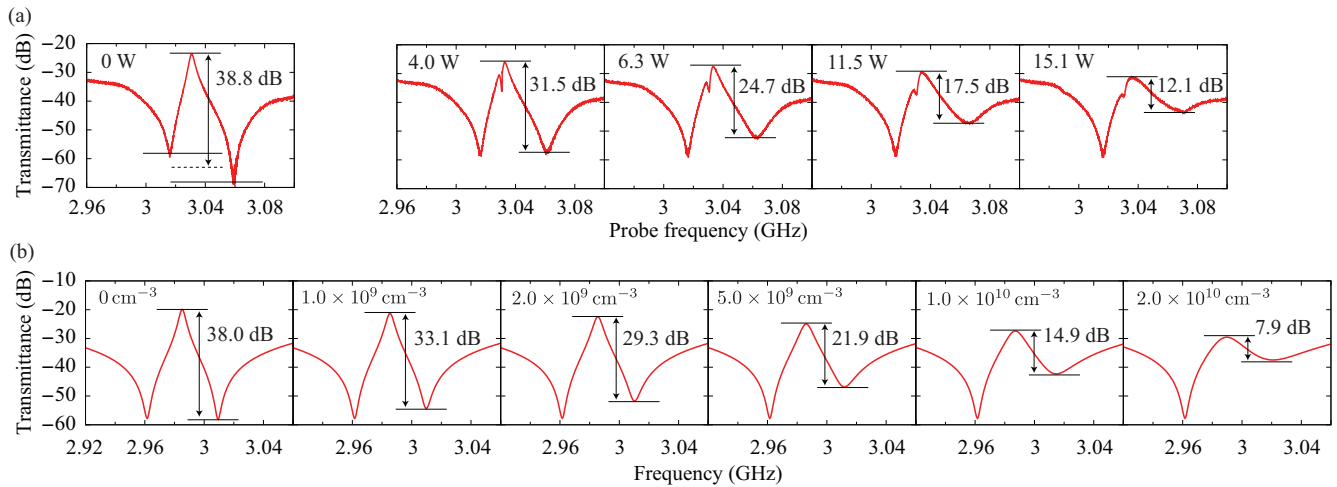}
  \caption{Comparison of (a) the experimentally measured transmission
  spectra for the probe wave for five different pump powers and (b) the numerically
  calculated transmission spectra for six different electron densities at
  the air pressure of $1.0\,\U{kPa}$. The numerical simulation is based on the
  assumptions that the plasma is uniformly generated in the gap of resonator H, that
  the electron temperature is 2\,eV, and that the elastic scattering
  cross section is $20 \times 10^{-16}\,\U{cm}^2$. Although the
  transmission window for $n\sub{e}=0\,\U{cm}^{-3}$ is symmetric, that
  for the pump power of 0\,W is slightly asymmetric
  because the direct coupling between resonators L and H is not completely
  canceled out in the experiment. Therefore, for simplicity, the value of $r\sub{t}$,
  which is defined in the main text, for the pump power of 0\,W is regarded as
  the ratio of the transmittance at the transmission peak frequency
  to the average of the transmittances at the transmission dip frequencies.}
  \label{fig:exp_sim}
 \end{center}
\end{figure*}

\begin{figure}[h]
 \begin{center}
  \includegraphics[scale=0.7]{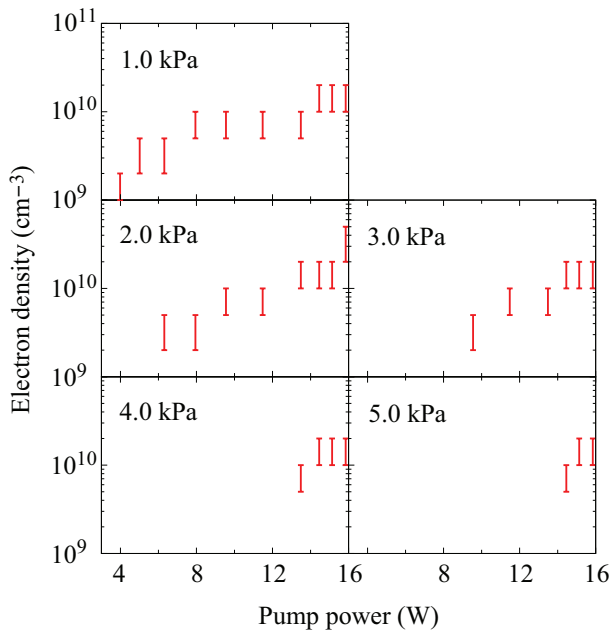}
  \caption{Electron density plotted against pump power for
  $p\sub{air}$ values ranging from 1.0\,kPa to 5.0\,kPa.
  The electron density was estimated to be in the range represented by the bars.
  The data points for $p\sub{air}=1.0\,\U{kPa}$, 3.0\,kPa, and 5.0\,kPa
  correspond to those in Fig.~\ref{fig:dip}. }
  \label{fig:dens}
 \end{center}
\end{figure}

Finally, the electron density of the microplasma is roughly estimated. The transmission spectrum of the metasurface was calculated using the commercial finite element software COMSOL Multiphysics, and the electron density was estimated by comparing the experimentally and numerically obtained values of the ratio $r\sub{t}$ of the transmittance at the transmission peak frequency to that at the higher-frequency transmission dip frequency. In this simulation, it was assumed that microplasma was generated uniformly in the gap of resonator H, that $T\sub{e} = 2\,\U{eV}$, and that $\sigma = 20 \times 10^{-16}\,\U{cm}^{2}$. The comparison of experimentally and numerically obtained transmission spectra for $p\sub{air} = 1.0\,\U{kPa}$ are shown in Fig.~\ref{fig:exp_sim}. Here the conductance of aluminum in the simulation, which was set to $3 \times 10^5\,\U{S/m}$, was determined such that $r\sub{t}$ for $n\sub{e}=0\,\U{cm}^{-3}$ in the simulation became closest to $r\sub{t}$ for the case without the pump wave in the experiment. The ratio $r\sub{t}$ decreases and the higher-frequency transmission dip shifts to higher frequencies with increasing the pump power (electron density) in the experiment (simulation). This implies that the experimental observation is qualitatively reproduced by the numerical simulation. In this study, if $r\sub{t}$ in the experiment is between $r\sub{t}$ for $n\sub{e}=n\sub{e1}$ and $n\sub{e2}$ in the simulation, the electron density in the experiment is estimated to be between $n\sub{e1}$ and $n\sub{e2}$. Figure \ref{fig:dens} shows the pump power dependence of the electron density for $p\sub{air}$ values ranging from 1.0\,kPa to 5.0\,kPa. The electron density increases with pump power and becomes of order $1\times 10^{10}\,\U{cm}^{-3}$ for a pump power of 15.8\,W. Note that the electron density was estimated based on very simple assumptions.

\section{conclusion}

The generation of microplasma in an EIT-like metasurface composed of radiatively coupled cut-wire resonators was investigated in this study. The threshold power for plasma ignition was lowest when the incident frequency coincided with the transmission peak frequency. This result is clear evidence that the local electric field enhancement is strongest at the transmission peak frequency, where the group index is maximized. The threshold power increased with $p\sub{air}$ ($\geq 1.0\,\U{kPa}$), indicating that the experimental conditions in this study lay on the right-hand side of the Paschen curve. A pump--probe measurement revealed that the transmission peak and one of the transmission dips were suppressed by the generation of microplasma. The center frequency of the suppressed transmission dip and the transmittance at that frequency increased with increasing the pump power, and the ratio of the transmittance variation to the frequency shift increased with $p\sub{air}$. This observation was qualitatively explained by the Drude dispersion relation of the permittivity of plasma. The electron density of microplasma was roughly estimated to be of order $1\times 10^{10}\,\U{cm}^{-3}$ for a pump power of 15.8\,W by comparing the measured and numerically calculated transmission spectra of the metasurface under assumptions that the plasma was uniformly generated in the gap of resonator H, that the electron temperature was 2\,eV, and that the elastic scattering cross section was $20 \times 10^{-16}\,\U{cm}^2$.

This study yielded two significant findings. First, it was demonstrated that microplasmas can be generated by relatively low-power microwaves in metasurfaces with a large group index. The threshold power for plasma ignition can be lowered if the metasurface structure is designed for a larger group index. This points to a convenient method of generating microplasmas with a high electron density. Second, it was shown that nonlinear and/or dynamic metasurfaces are realized using microplasmas. Even a small change in the properties of a microplasma causes a large change in the electromagnetic response of the metasurface, especially near its resonance frequency. This implies that metasurfaces with plasmas can be used for advanced control of electromagnetic waves, as described in previous studies.\cite{sakai12_psst,iwai15_apex,iwai15_pre,kourtzanidis16} The integration of studies on microplasmas and metasurfaces would contribute to the development of plasma physics, electromagnetic optics, and other related fields of study.

\begin{acknowledgments}
This research was supported in part by \mbox{JSPS} \mbox{KAKENHI} Grant No. \mbox{JP16K14249} and by a research granted from The Murata Science Foundation.
\end{acknowledgments}

%

\end{document}